\documentclass[preprint, prd]{revtex4}
\usepackage{amsmath}    
\usepackage{graphicx}   
\usepackage{verbatim}   
\usepackage{color}      
\usepackage{subfigure}  
\usepackage{hyperref}   
\usepackage{epsfig}
\usepackage{graphicx}
\usepackage{color}
\usepackage{amsmath, amsthm}

\definecolor{lred}{rgb}{1,0.3,0.3}
\definecolor{red}{rgb}{0.5,0,0}
\definecolor{dblue}{rgb}{0,0,.6}

\begin{document}

\newcommand{\D}{\displaystyle} 
\newcommand{\T}{\textstyle} 
\newcommand{\SC}{\scriptstyle} 
\newcommand{\SSC}{\scriptscriptstyle} 

\newcommand{\be}{\begin{eqnarray}}
\newcommand{\ben}{\begin{eqnarray}\nonumber}
\newcommand{\ee}{\end{eqnarray}}
\newcommand{\nee}{\nonumber \end{eqnarray}} 
\newcommand{\nn}{\nonumber \\} 
\def\AJ{{ Astron. J.} }
\def\ARAA{{ Ann. Rev. Astron. \& Astrophys.} }
\def\ApJ{{ Astrophys. J.} }
\def\ApJL{{ Astrophys. J. Letters} }
\def\ApJS{{ Astrophys. J. Suppl.} }
\def\ApP{{ Astropart. Phys.} }
\def\AA{{ Astron. \& Astroph.} }
\def\AAR{{ Astron. \& Astroph. Rev.} }
\def\AAL{{ Astron. \& Astroph. Letters} }
\def\AASu{{ Astron. \& Astroph. Suppl.} }
\def\AN{{ Astron. Nachr.} }
\def\IJMP{{ Int. J. of Mod. Phys.} }
\def\JCAP{{\it Journ. of Cosmol. \& Astropart. Phys.}Ê}
\def\JGR{{ Journ. of Geophys. Res.}}
\def\JHEP{{ Journ. of High En. Phys.} }
\def\JPhG{{ Journ. of Physics} {\bf G} }
\def\MNRAS{{ Month. Not. Roy. Astr. Soc.} }
\def\Nature{{ Nature} }
\def\NewAR{{ New Astron. Rev.} }
\def\PASP{{ Publ. Astron. Soc. Pac.} }
\def\PhFl{{ Phys. of Fluids} }
\def\PLB{{ Phys. Lett.}{\bf B} }
\def\PR{{ Phys. Rev.} }
\def\PRD{{ Phys. Rev.} {\bf D} }
\def\PRL{{ Phys. Rev. Letters} }
\def\RMP{{ Rev. Mod. Phys.} }
\def\Science{{ Science} }
\def\ZfA{{ Zeitschr. f{\"u}r Astrophys.} }
\def\ZfN{{ Zeitschr. f{\"u}r Naturforsch.} }
\def\etal{{ et al.}}

\hyphenation{mono-chro-matic sour-ces Wein-berg
chang-es Strah-lung dis-tri-bu-tion com-po-si-tion elec-tro-mag-ne-tic
ex-tra-galactic ap-prox-i-ma-tion nu-cle-o-syn-the-sis re-spec-tive-ly
su-per-nova su-per-novae su-per-nova-shocks con-vec-tive down-wards
es-ti-ma-ted frag-ments grav-i-ta-tion-al-ly el-e-ments me-di-um
ob-ser-va-tions tur-bul-ence sec-ond-ary in-ter-action
in-ter-stellar spall-ation ar-gu-ment de-pen-dence sig-nif-i-cant-ly
in-flu-enc-ed par-ti-cle sim-plic-i-ty nu-cle-ar smash-es iso-topes
in-ject-ed in-di-vid-u-al nor-mal-iza-tion lon-ger con-stant
sta-tion-ary sta-tion-ar-i-ty spec-trum pro-por-tion-al cos-mic
re-turn ob-ser-va-tion-al es-ti-mate switch-over grav-i-ta-tion-al
super-galactic com-po-nent com-po-nents prob-a-bly cos-mo-log-ical-ly
Kron-berg Berk-huij-sen}
\def\simle{\lower 2pt \hbox {$\buildrel < \over {\scriptstyle \sim }$}}
\def\simge{\lower 2pt \hbox {$\buildrel > \over {\scriptstyle \sim }$}}
\def\intunits{{\rm s}^{-1}\,{\rm sr}^{-1} {\rm cm}^{-2}}


\title 
{A supersymmetric model for triggering Supernova Ia in isolated white dwarfs}
\author
{Peter L. Biermann}
\affiliation{Dept.\ of Physics \& Astronomy, Univ.\ of Alabama, Tuscaloosa, AL}
\altaffiliation{[PLB also at] Max-Planck-Institute for Radioastronomy, Bonn, Germany}
\altaffiliation{FZ Karlsruhe, and Physics Dept., Univ.\  Karlsruhe, Germany}
\altaffiliation{Dept.\ of Physics \& Astronomy, Univ.\ Bonn, Germany}
\altaffiliation{Dept. of Physics, Univ.\ of Alabama at Huntsville, AL}
\author
{L. Clavelli}
\affiliation{Dept.\ of Physics \& Astronomy, Univ.\ of Alabama, Tuscaloosa, AL}
\date{\today}

%
%

\begin{abstract}
We propose a model for supernovae Ia explosions based on a phase transition to a supersymmetric state which becomes the active trigger for the deflagration starting the explosion in an isolated sub-Chandrasekhar white dwarf star.
With two free parameters we fit the rate and several properties of type Ia supernovae and address the gap in the supermassive black hole mass distribution.  One parameter is a critical density fit to about $3 \cdot 10^7$ g/cc while the other has the units of a space time volume and is found to be of order $0.05\,$ Gyr  $R_E^3$ where $R_E$ is the earth radius. The model involves a phase transition to an exact supersymmetry in a small core of a dense star.

\end{abstract}

\maketitle

\section{Introduction}
   It is central to string landscape ideas that the universe can exist in a (large) number of local minima in an effective potential and makes transitions between them.  One of these is perhaps a local minimum with a very large vacuum energy describing the inflationary era.  A second one is our broken supersymmetric (susy) world described perhaps as the Minimal Supersymmetric Standard Model (MSSM) with soft breaking parameters lifting the masses of the susy partners to hundreds of GeV and with a small positive vacuum energy.  It is likely that the world of exact susy described by the MSSM with soft breaking parameters set to zero and vanishing vacuum energy also exists.  In this world the quarks and leptons would have degenerate spin zero partners as would their baryon and meson bound states.  Such an exactly supersymmetric universe could be the ground state of the string landscape or at least a long lived intermediate state with the true ground state being a state of negative vacuum energy as in the conformal field theory/Anti-deSitter correspondence (CFT/AdS). 
The transition between our universe and the zero vacuum energy state is describable as a decay of the false vacuum.
The transition probability would presumably be affected by the presence of matter and we assume that it is enhanced rather than suppressed as suggested by analyses in lower dimensional models and by the fact that the vacuum decay probability is an increasing function of the energy density difference between the two states.

Supernova Ia (SN Ia) explosions have been successfully used to provide strong evidence that the expansion of the universe is accelerating.  Vital to the argument is the understanding of the calibration of the light curves, and for that we need insight into the explosion mechanism.  The primary standard model picture for SN Ia prior to 2010 was the single degenerate model \cite{HillebrandtN00} whereby a white dwarf accretes matter from a main sequence binary partner until it approaches the Chandrasekhar limit and collapses.  This model has been greatly disfavored by recent x-ray analysis \cite{Gilfanov}.  Although many doubly degenerate models whereby two white dwarfs merge or collide are perhaps still viable, such models involve a host of free parameters with, consequently, little predictive power.  It is not known how many suitable binary systems exist or what would happen to the many systems without the fine-tuned orbital parameters needed to reproduce observations.
More complicated, shrouded, singly degenerate accretion models could also be devised to circumvent the result of \cite{Gilfanov}, but these add free parameters and extra complexity.  

In view of the doubts thrown into the debate of what SN Ia really are, the time may be ripe to reconsider phase transition models in parallel with further exploration of standard model possibilities.   In a recent review \cite{Roepke}, sub-Chandrasekhar explosions, of which our current model is an example, are identified as one of two scenarios which might ultimately explain the majority of SN Ia events.  

%

The primary features of SN Ia which we address in our two parameter model
based on string landscape and vacuum decay ideas are:
\begin{enumerate}
\item{no hydrogen in emission;}
\item{homogeneity;}
\item{light curve dominated by $^{56}Ni$ production and decay;}
\item{rate: About one per century per galaxy\\
(more in galaxies with enhanced star production but also observed in old galaxies);}
\item{some evidence for appreciable departures from spherical symmetry;}
\end{enumerate}

Item 1 is naturally fit if the SN Ia progenitors are white dwarf stars with
no circumstellar hydrogen as occurs both in our model and in doubly degenerate accretion models. 

As discussed below,
our model predicts a narrow range of progenitor white dwarf masses which 
naturally results in the SN Ia homogeneity (Item 2 above) with, however, some spread in energy release. 

The supersymmetric (susy) bubble grows in the present model from a sub-atomic size to a
macroscopic size which, however, is limited by the lack of degeneracy pressure 
in the susy phase.  The energy release induces fusion in the outside normal phase leading to the observed $^{56}Ni$ production which is a natural endpoint
of fusion in a  carbon-oxygen white dwarf (Item 3).

The rate of SN Ia is easily fit in our model using the known mass spectrum of 
white dwarf stars (Item 4).  The observed rate yields a strong relation between the model's two free parameters.   This feeds back into the narrowness prediction and
also naturally yields the lifetime observations of item 4.

Although the nucleation of the susy bubble begins most probably at the stellar center, the average location of the nucleation point is quantitatively predicted to be off-center leading naturally to an off-center explosion (item 5).  
This average location is predicted to be
just inside the sphere containing the entire region above critical density with a broad distribution throughout this region. 

A fit to one of our two free parameters suggests  
that the supersymmetric (susy) phase transition described above occurs in single white dwarfs of mass from about 0.9 to about 1.3 solar masses, starting a deflagration front, which then triggers the explosion as a supernova \cite{Mazzali2007,HillebrandtN00}.
Since the data clearly show that Supernovae Ia occur both in stellar systems, which are constituted out of mostly old stars (elliptical galaxies), and in stellar systems made of old and many new stars (late Hubble type spirals) \cite{Pritchet2008}, there has to be a very broad delay time distribution from making a white dwarf to its explosion.  We emphasize that what we call here a delay time is more akin to the decay of a radioactive nucleus, so any particular white dwarf passing the threshold may still have a chance to go susy somewhat earlier and also somewhat later than the nominal delay time.  For single stars a broad delay time distribution between creation and explosion constitutes a challenge, but a susy phase transition offers such a prediction.
Super-massive black holes have been observed over a wide mass range, from about $3 \cdot 10^{6} \, M_{\odot}$ to $3 \cdot 10^{9} \, M_{\odot}$, with some outliers to lower masses such as $3 \cdot 10^{5}\, M_{\odot}$ \cite{Barth2005}.
There is also strong evidence for stellar black holes, in the approximate mass range 3 - 5 $M_{\odot}$.  However, there is currently no convincing evidence for a large number of black holes in the intermediate mass range 
\cite{Caramete}.  These are either difficult to detect, or just may not exist. Observational arguments for the existence of a few do exist \cite{Irwin2010}; expected are many.  Here we pursue the notion, that such black holes can derive from the agglomeration of very massive stars: we explore the concept that across the mass range of the gap the final collapse of the star leads to a susy phase transition in the core.  This leads to an additional energy input possibly pushing these stars into an enhanced mass loss during collapse, and may result in a final compact object of much less than $10^{5} \, M_{\odot}$.

From many points of view, the theory of violent astrophysical events would be less problematic if there were a new source of energy release beyond the standard model.  For instance, in the standard model the mass of a white dwarf grows by accretion
to the Chandrasekhar limit.  However a star near this limit has a binding energy greater than available from fusion reactions making it impossible to totally unbind the star without borrowing energy from the gravitational energy
released during the accretion.  This energy must be repaid as the star explosively expands making the timing critically sensitive. 
Similarly  
in earlier years it was thought that the energy required to blast off the outer shell of massive stars in a Type II supernova was provided by neutrinos but many detailed studies have found that such supernovae would stall because of insufficient numbers of neutrinos and the weakness of neutrino interactions \cite{Barwick}. 
 In the case of the Type Ib/c supernovae of very massive stars magnetic fields may help cause the explosion \cite{GBK-SN,CRIV-CREAM}.
 
A phase transition to exact susy in dense stars could be nature's way to release the energy stored in a Pauli tower of fermions.  
A susy phase transition in white dwarf stars (WD's) could eliminate the need for accretion and allow an isolated white dwarf to explode with the required extra energy input and a correctly predicted rate.

\section{ The Decay Assumption}

In the vacuum the probability per unit time per unit volume for the decay of the false vacuum is governed by the Coleman-DeLuccia formula \cite{Coleman}.

\be
     \frac{d^2P}{dt d^3 r} \; = \; A_C e^{-B(vac)}  
\label{AemB}
\ee

\noindent with

\be
B(vac) \;  = \; \frac{27 \pi^2 S^4}{2\,\hbar\,c\,\epsilon^{3} }
\label{AemB2}
\ee
$A_C^{-1} \, = \tau_0 \, V_0$ is a parameter with the units of a space-time volume.  $S$ is the surface tension of the true vacuum bubble in the dominantly false vacuum background.  In the case of a susy phase transition $\epsilon$ is the difference between the vacuum energy density of our universe and the zero vacuum energy density of an exactly supersymmetric background.

In dense matter it is expected that the transition is accelerated since the energy density difference between a broken susy ground state and that of an exact susy ground state with the same additive quantum numbers (baryon and lepton numbers) is dominated by the energy stored in the Pauli towers.  Thus it is plausible that the phase transition formula could, to at least a first approximation, be given by replacing $\epsilon$ with the total energy density difference $\epsilon + \Delta \rho\,c^2$.  At present however, a rigorous proof of the acceleration exists only in lower dimensions \cite{Gorsky}.  
 
Other attempts to model
matter induced or catalyzed vacuum decay, although different from the present proposal, can be found in ref.\,\cite{inducedvacdecay}.
 
In exact susy, the degeneracy of bosons and fermions plus the availability of a conversion mechanism \cite{CK05,growth} 
from a pair of fermions to a pair of partner scalars implies that $\Delta \rho \, c^2$ is the Pauli excitation energy density of the fermions. The Fermi gas model predicts for this excitation energy in a heavy nucleus of $N$ neutrons and $Z$ protons
\ben
   \Delta \rho = \rho \frac{\Delta E}{A\,M_n\,c^2} =  \frac{1}{2}\left( (\frac{2\,N}{A})^{5/3} + (\frac{2\,Z}{A})^{5/3} \right)\\
\cdot \frac{3\,(9\pi)^{2/3}}{40}\frac{\hbar\,\rho}{M_n\,c\,R_0} \approx 0.02 \rho\quad .
\label{deltarho}
\ee

Here $M_n$ is the nucleon mass, $R_0$ is the nuclear constant $1.2$ fm, and the approximate value is given for a nucleus with $N=Z=A/2$.  This energy release is about three times that of standard hydrogen fusion and some twenty times greater than the  carbon or oxygen fusion which could provide the energy release in a standard model SN Ia.

Thus the parameter controlling the exponential factor in the transition rate would then be

\be
      B = \frac{27 \pi^2 S^4}{2\,\hbar\,c\,(\epsilon+\Delta\rho\,c^2)^3 } \quad .
\label{Bmatter}
\ee

We would then have
\be
     \frac{d^2P}{dt d^3 r} \; = \;\frac{1}{\tau_0 V_0}\,e^{-B} \quad . 
\label{transprob}
\ee

This equation should be taken as the basic assumption of this paper with the previous discussion of vacuum decay and possible connections with string landscape ideas providing some motivation and inspiration.
Although the vacuum decay theory, based as it is on the WKB approximation, is subject to corrections,  
the current paper attempts to explore the consequences of Eq.(\ref{transprob}) as written with $\tau_0 V_0$ and $S$ treated as two free parameters.
We do not provide a theory of susy breaking although, when one is eventually found, it should predict the
values of our free parameters.  We use the observation that the vacuum energy of our broken susy universe is positive while, in exact flat space susy, the vacuum energy vanishes.  An auxiliary assumption is that the degenerate mass of fermions and bosons is equal to the fermion mass in the broken susy world.  This would be as in the MSSM or one of its extensions with the soft susy breaking parameters set to zero.

With our two parameters we fit the SN Ia rate and explain the energy release in SN Ia (approximately) as well as the narrowness of 
the progenitor mass distribution and the time delay distribution from the white dwarf birth i.e. the onset of degeneracy.
Note that white dwarfs in our predicted mass range do not spend appreciable time in the red giant phase.

In dense matter containing heavy nuclei (above helium in which all nucleons are in the lowest level), $\epsilon$ is negligible compared to $\Delta \rho\,c^2$.
  
We would therefore write, using the Fermi gas model and replacing $S$ in terms of a critical density $\rho_c$,
\be
     \frac{d^2P}{dt d^3 r} \; =\; \frac{1}{\tau_0 V_0}\,e^{-{(\frac{\rho_c}{\rho})^3}} \quad .
\label{transprob2}
\ee

\begin{figure}[ht]
\centering
\includegraphics[scale=0.65]{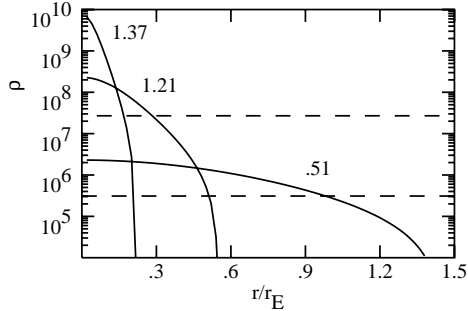}
\caption{\baselineskip=18pt
The density of three white dwarfs as a function of radius relative to the Earth radius $r_E$.  Dashed lines give the range of previously discussed critical densities.  Masses of the three white dwarf examples are given in $M_{\odot}$.  These Chandrasekhar density distributions go into the calculation of $V_c$ in Eq.\,(\ref{transprob3}).}
\label{D8cgsg} 
\end{figure}

The white dwarf density, $\rho$, as a function of radius is given as in fig.\,\ref{D8cgsg} depending on the total mass of the star.  
The phase transition probability per unit time per unit volume, Eq.\,(\ref{transprob2}), increases rapidly with $\rho$ until 
$\rho$ becomes of order $\rho_c$ at which point it saturates.  For denser media, the transition rate is proportional to the volume.  Thus, if we choose $\rho_c$ as near the white dwarf density, the transition rate in neutron stars will be at least eight orders of magnitude lower than that in white dwarfs.  If there was a degenerate quark gluon plasma in the very early universe, one could consider whether the universe might have turned supersymmetric at that stage thus ruling out our model but, in fact, the space time volume of this phase would have been too small compared to $\tau_0 V_0$ to effect the phase transition with an appreciable probability.  We would estimate the probability as roughly
$10^{-10}$.  

For present purposes we will ignore the additional Pauli energy stored in electronic states.   

\section{Growth of the susy Core}

In dense matter the critical radius, generalized from the vacuum decay result would be

\be
     R_c =  \frac{3 S}{\epsilon+\Delta\rho\,c^2}\quad .
\label{criticalradius}
\ee

Bubbles nucleated with less than this critical radius will be immediately quenched.  This follows from the vacuum decay result if, in dense matter, the relevant quantity is the total energy density difference between the two states and not just the vacuum energy difference as in the basic assumption of Eq.\,(\ref{transprob}).
A bubble of critical (or greater) radius will grow as long as its radius exceeds the density dependent critical radius of Eq.\,(\ref{criticalradius}).  
Replacing $S$ by a critical density as discussed above leads to a critical radius of sub-atomic size in a white dwarf but still much larger than a nuclear radius.
In a medium of uniform density (such as the vacuum) the bubble will grow without limit but a susy bubble nucleated within a dense star will not grow beyond the boundary of the star since outside the star the critical radius is of galactic scale.
However, there is also a hydrodynamic limit on the growth of the bubble in dense matter due to the absence of degeneracy pressure.  The net effect is that the susy core never grows beyond a small fraction of the host star.     

It is important to work out the susy phase transition front growth:  A phase transition with energy input can be seen as analogous to an HII region ionization front; the ionization injects energy and there is a surface discontinuity, usually a shock front.  Writing for the speed of sound ahead of the front, in the normal white dwarf matter, $C_\mathrm{wd}$, and for the speed of sound behind it, in the susy matter $C_\mathrm{susy}$, the advance speed $U_D$ is (\cite{Spitzer}) given by what is called a D-type front:

\be
U_D \; \simeq \; \frac{C_{\mathrm WD}^2}{2 \, C_{susy}}\quad.
\label{susyfrontD}
\ee

Since very likely $C_{\mathrm WD}/C_\mathrm{susy} < 1$, it is probable that the advance speed is below the speed of sound, and so the susy core grows only slowly.  In lower density regions, the speed of sound is considerably less so the bubble growth is further slowed.
Particles that are light in both phases can pass freely through the phase boundary and induce normal fusion in the region outside the core.
Normal fusion reactions outside of the susy core produce the Nickel that governs the SN Ia light curves.  Since, according to calculations (see \cite{Stehle}),  the Nickel seems to be produced at intermediate $r$, it is an important
feature of our model that the susy core remains small compared to the radius of the star \cite{growth}.

From Eq.\,(\ref{transprob2}) we see that each star of mass $M$ has a characteristic
lifetime $\tau(M)$ defined by
\be
     \frac{dP}{dt} = \frac{1}{\tau_0\,V_0}  \int d^3r\, e^{-(\rho_c/\rho(r))^3} = \frac{1}{\tau_0}\frac{V_c}{V_0}\, \equiv \, \frac{1}{\tau(M)} \quad  .
\label{transprob3}
\ee

For very dense stars $V_c$ is approximately the volume of the star at or above the critical density.  To consider the transition in a dense helium star
one would have to replace Eq.\,(\ref{deltarho}) by the appropriate formula for an electronic rather than nuclear degeneracy.  For a given susy bubble surface tension, the transition probability per unit time per unit volume would then be much less than for a white dwarf star.
In the absence of new star production this would lead to a number $N(M,t)$ surviving a time $t$ after achieving a constant density defined by
\be
     N(M,t) = N(M,0)\,e^{-t/\tau(M)}
\label{decaylaw}
\ee 
 
The complement of the implied probability defines the likelihood that
a white dwarf of mass $M$ will explode in time less than $t$. 
The current model can be ruled out if a significant sample of white dwarfs
can be found with ages long compared to their $\tau(M)$ defined in Eq.\,(\ref{transprob3}).  At present only
low mass white dwarfs are known to have existed over gigayear time scales. 
New white dwarfs will be produced at a declining rate as the raw material in the galaxy is depleted.  We can, for simplicity, parametrize this in terms of a star formation time scale, $\tau_\psi$,
and a normalization, $A(M)$:
\be
     \frac{dN(M,t)}{dt} = \frac{A(M)}{\tau_\psi} e^{-t/\tau_\psi}\quad .
\ee
If we combine Eq.\,(\ref{transprob3}) with this rate of production of new white dwarfs we would predict a time dependent number of white dwarfs given by
\be
   N(M,t) = A(M) \frac{\tau(M)}{(\tau(M)-\tau_\psi)}\,\left(e^{-t/\tau(M)} - e^{-t/\tau_\psi}\right)
\ee
%

Equation $(\ref{transprob3})$ defines a natural lifetime $\tau(M)$ for any broken susy object
containing heavy nuclei, but for objects far below the white dwarf density, such as our current sun, this lifetime is, by our choice of the $\tau_0 V_0$ parameter, effectively infinite, i.e. many orders of magnitude longer than the current age of the universe.

Just as it is statistically possible for an elementary particle to decay in a time less than its natural lifetime, a white dwarf in our model could also produce a supernova in a time less than $\tau(M)$ although $\tau(M)$ would be the average life of an ensemble of such stars. 
The probability for a phase transition to occur at some time $>t$ after reaching its classical steady state would be
$e^{-t/\tau(M)}$.  Note that for high mass white dwarfs ($M > 0.9\,M_\odot$) the time spent in the red giant and previous phases is small compared to a gigayear.

It has been suggested \cite{Frampton}
that the critical radius in vacuo should be the galactic radius ($\approx 4.7\,\cdot 10^{22}\,\mathrm{cm}$).  This would lead to a surface tension $S=8.9\,\cdot 10^{13}\,\mathrm{erg/cm}^2$ and then to $\rho_c = 3\,\cdot 10^5\,\mathrm{g/cc}$.  An independent estimate from applying the susy phase transition model to gamma ray bursts \cite{CK05} leads to comparable values from $\rho_c \approx 10^6\,\mathrm{g/cc}$ to $\rho_c \approx 3\,\cdot 10^7\,\mathrm{g/cc}$.  Thus from two very different starting points a critical density near the white dwarf density is suggested.  As we shall see, the edge of the high mass black hole distribution also suggests interesting new physics near this density.

Without loss of generality one can choose $V_0$ to be the maximum $V_c$ for all white dwarf masses at a given $\rho_c$.
Then the parameter $\tau_0$ represents, for given $\rho_c$, the minimum quantum mechanical lifetime of white dwarfs against nucleation of a susy core. The lifetime of any other star is inversely proportional to its critical volume, $V_c$.
Ultimately, a star with such a susy core must either be totally disrupted by the susy energy release or collapse into a black hole due to the absence of degeneracy pressure.  The time over which a star can survive with a susy core is dependent on the rate of susy energy release.

\begin{figure}[ht]
\centering
\includegraphics[scale=0.65]{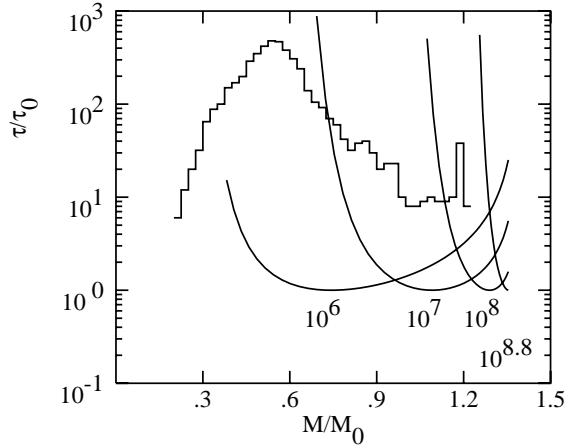}
\caption{\baselineskip=18pt The relative delay time distribution, $\tau/\tau_0$, as a function of mass for critical densities of $10^6$, $10^7$, $10^8$, and $6.3 \cdot 10^8$  g/cc.  For an approprate choice of $\rho_c$, the narrowness of the SN Ia distribution is naturally explained.  We superimpose the
histogram of the observed mass distribution of white dwarfs above an observed temperature of 12,000 K \cite{Madej}. For $\rho_c > 10^7$ g/cc and $\tau_0 = 0.5$ Gyr only high mass white dwarfs undergo the susy transition with a
lifetime less than the current age of the universe.}
\label{SNgd}
\end{figure}

\begin{figure}[ht]
\centering
\includegraphics[scale=0.65]{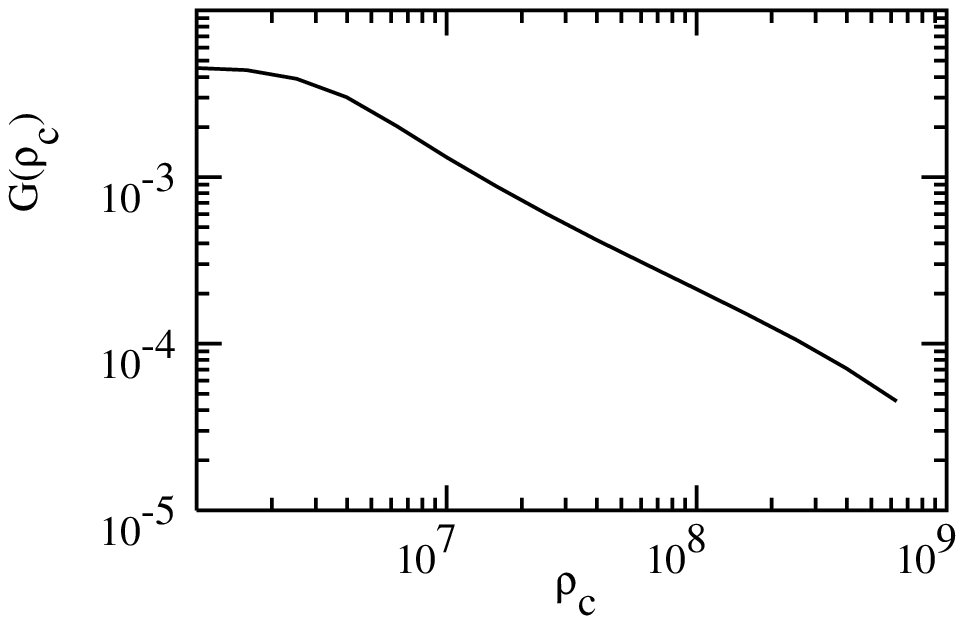}
\caption{\baselineskip=18pt The function $G(\rho_c)$ which gives via Eq.\,(\ref{SNIarate}) the supernova rate.
$\rho_c$ is given in g/cc.  }
\label{Rhogd} 
\end{figure}

\begin{figure}[ht]
\centering
\includegraphics[scale=0.5]{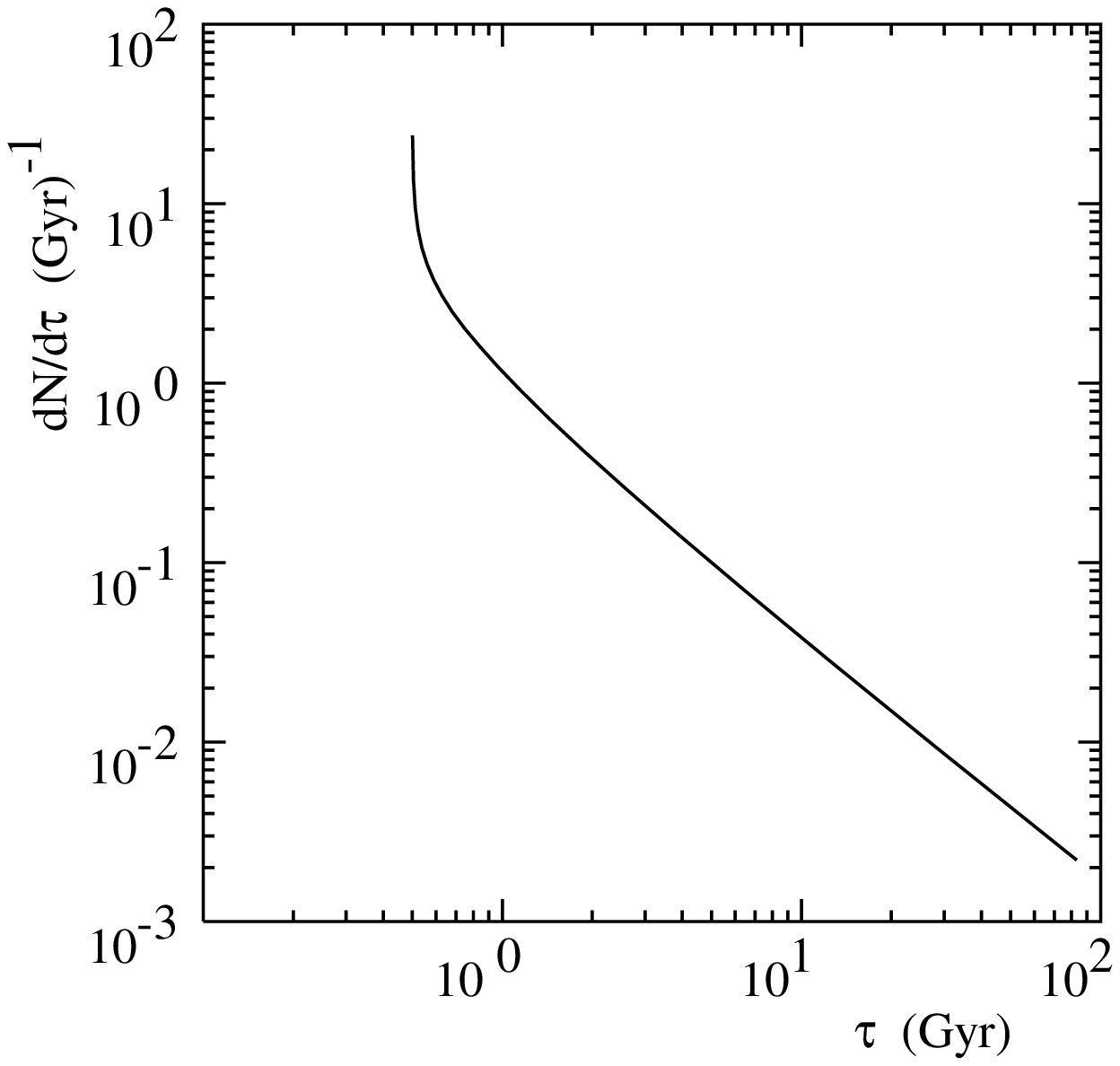}
\caption{\baselineskip=18pt The relative lifetime distributions for the critical density of $3.55 \cdot 10^{7}$ g/cc, as a function of $\tau$ for $\tau_0=0.5$ Gyr.  The evident
scaling behavior in Eq.\,(\ref{dNdtau}) shows that the graph is a function of
$\tau/\tau_0$ only.
The distribution is sharply peaked at $\tau = \tau_0$.}
\label{dndtaufig}
\end{figure}

In a single star model for SN Ia we obviously have no need to push a star's mass up with time via accretion from a binary partner. What we propose is commonly referred to as as a sub-Chandrasekhar model: the explosion of a white dwarf below the Chandrasekhar mass limit.  In our model this is triggered by the phase transition to a susy phase inside a very small core of the white dwarf.  Thus this model makes no distinction between C+O white dwarfs and Ne-O white dwarfs, except in so far as they exist in different white dwarf mass ranges.

The explosion rate can then be written as
\be
       \frac{dN_{SNIa}}{dt} = N_{\mathrm WD} \frac{G(\rho_c)}{\tau_0}
\label{SNIarate}
\ee
where
\be
        G(\rho_c) = \int dM \frac{1}{N_\mathrm{WD}}\frac{dN_{\mathrm WD}}{dM}  \frac{V_c}{V_0} \quad .
\label{SNIarateG}        
\ee

Here the white dwarf mass distribution - which can be modelled over the relevant mass range using the Salpeter mass function and the lifetime of stars on the main sequence - is taken to follow the high mass tail of the Sloan white dwarf sample of 4621 stars with $T_{eff} > 12000 K$ \cite{Madej,CNMM07}. 
The suggested white dwarf mass distribution in the high mass tail going into Eq.\,(\ref{dNdtau}) varies as $M^{-4.78}$ although the result is not greatly sensitive to the exact behavior since the relevant mass range is small.

 $N_\mathrm{WD}$ in Eq.\,(\ref{SNIarate}) is the total number of white dwarfs in the Milky Way galaxy, estimated at near $10^{10}$ 
\cite{Liebert2005}.   
Statistics suggest \cite{Mannucci2005} that SN Ia are about as common as SNe from higher masses.  Assuming  that all white dwarfs above some particular zero age main sequence (ZAMS) mass and below $9 \; M_{\odot}$ become SN Ia, and that all stars above ZAMS mass $9 \; M_{\odot}$ become other kinds of SNe, suggests \cite{Salpeter1955} that this critical mass is about $5.5 \; M_{\odot}$, corresponding to a white dwarf mass of about $0.9 \; M_{\odot}$ again, approximately.  This is the mass at which the delay time is still less than the Hubble time, so that we still can get a few white dwarfs exploding at that mass within the age of a Galaxy.  

The SN Ia rate is found to be a function of both the total number of
white dwarf stars in a galaxy and the rate of star production
\cite{ScannapiecoBildsten}:
\be
     \frac{dSNIa}{dt} = A \frac{M_{\mathrm gal}}{10^{10} M_\odot} + B \frac{\frac{dM_{\mathrm gal}}{dt}}{10^{10}M_\odot/{\mathrm Gyr}}
\label{Scannapieco}
\ee
with $A=0.04_{0.011}^{+0.016}$ and $B=2.6 \pm 1.1$.\\

The observation that the supernova rate is greatly enhanced in galaxies with elevated star production rates suggests that there is a component of the supernova rate at small delay times.  The observation that supernovae continue to be produced in older galaxies suggests that there
is also a component with longer delay times.  Standard model approaches to SNIa strive to be compatible with this observation leading perhaps to a bi-modal delay time hypothesis but no firm prediction.

In our model there is a 
natural spike in the lifetime distribution near some minimum $\tau_0$ 
plus a tail to much higher lifetimes.
This distribution is given by
\be
     \frac{dN}{d\tau} = \int dM \frac{dN_{\mathrm WD}}{dM}\;\delta(\tau-\tau(M))
\label{dNdtau}
\ee

As can be seen from fig.\;\ref{SNgd} $\tau(M)$ is approximately
parabolic on a log-linear scale.  This implies that near $\tau=\tau_0$ there is an integrable singularity:
\be
     \frac{dN}{d\tau} \approx \frac{1}{\tau\,\sqrt{\ln{\tau/\tau_0}}}
\ee

Without relying on this approximate behavior the numerical solution of Eq.\,(\ref{dNdtau}) is given in fig.\,\ref{dndtaufig} showing clearly the spike at low lifetimes.  On the y axis in fig.\,\ref{dndtaufig} is the predicted number of white dwarfs with given lifetime in a sample of the size of the Sloan white dwarf sample of 4621 stars \cite{Madej,CNMM07}. 
The logarithmic spike predicted in the present model is similar to and
probably observationally indistinguishable from the empirical fit
\be
     \frac{dN}{d\tau} \approx \frac{1}{\sqrt{\tau}}
\ee
suggested in ref.\,\cite{Pritchet2008}.
These estimates suggest a critical density of order $3 \cdot 10^{7}$ g/cc, which we have used in fig.\,\ref{dndtaufig}.
The instantaneous SN Ia rate as given in Eq.\,(\ref{SNIarate}) is a function only of the total number of white dwarfs in the sensitive mass range but, during periods of elevated star production rate, this number is also temporarily elevated leading to the $B$ term in Eq.\,(\ref{Scannapieco}).  As the star production rate declines a dip should appear in the white dwarf mass distribution at the mass of minimum lifetime although the statistics are
too meager at present to see this in our galaxy.  Another prediction of
the current model with our fitted value of $\rho_c$ is that the total
mass ejected in the supernova should range from $1.0\,M_\odot$ to the Chandrasekhar mass.  In a binary accretion model the total mass ejected 
should be very close to or above the Chandrasekhar mass.  Current measurements are not sufficiently precise to rule out or confirm either model.

We can check how much energy is released by the phase transition of a very small part of the white dwarf:  If we require that
the susy energy release plus that from fusion outside the susy core is equal to the binding energy of the outside shell plus some $10^{51}$ ergs of kinetic energy we obtain that the susy core is very much smaller than the stellar radius and the proportional mass in Nickel can easily vary by over an order of magnitude.  
It is also possible that the variability in the distance of the susy nucleation point from the center as shown
below in fig.\,\ref{nucleationpoint} contributes to the variability in
the amount of Nickel produced.  

For SN Ia cores we need less than $10^{-3.5} \, M_{\odot}$ (\cite{Mazzali2007}) for an initiating phase transition.  Since the susy phase transition provides more energy than even hydrogen burning, even less matter than in normal models is required to initiate the deflagration.

One proof of our approach could be if these very small compact susy objects could be found remaining after the explosion.  In some respects these objects might resemble black holes of anomalously small mass.

Folding the white dwarf mass function with the delay time distribution allows on the one hand to obtain the SN Ia progenitor mass distribution (see fig.\,\ref{SNgd}), and also to obtain the SN Ia distribution as a function of delay time 
(fig.\,\ref{dndtaufig}).  Most SN Ia are produced after a lifetime near $\tau_0$ although
there is a tail extending to higher lifetimes (see fig.\,\ref{dndtaufig}). 
With a small $\tau_0$ we naturally understand the greater rate of SN Ia in
galaxies with an elevated star production rate and the lower rate in our galaxy.

As greater statistics of
high mass white dwarfs are accumulated, we would predict a dip in the white dwarf mass distribution at the minimum of the $\tau/\tau_0$ curve for chosen critical density in fig.\,\ref{SNgd}.
 Since our approach yields a very narrow mass distribution for those white dwarfs that explode as SN Ia, the energy distribution is also very tight and so a use as a standard candle is justified.  The apparent luminosity is, therefore, a good measure of its distance.

Data also show \cite{Raskin09} a relatively large number of SN Ia with a delay time of order $5 \cdot 10^{8}$ yrs.  Combining these two points yields the critical density and the critical time scale.  Therefore a clear prediction of this model is a relatively sharp sub-Chandrasekhar mass distribution (see, e.g., ref. \,\cite{KJTCBT2010}), peaking at the minimum of the delay time distribution.  Thus our estimates for the key parameters are $\rho_c \, \simeq \, 3 \cdot 10^{7}$ g/cc, $\tau_0 \, \simeq \, 5 \cdot 10^{8}$ yrs, resulting in a typical progenitor mass near $1.2 \, M_{\odot}$, mostly an original C-O composition, and an average SN Ia rate per galaxy of about 1/100 yrs, see Eq.\,(\ref{SNIarate}) and the accompanying fig.\,\ref{Rhogd}.  Considering the range of uncertainty in the observations (\cite{Mannucci2005,Raskin09}), the time scale could be yet shorter, implying a smaller value of $G(\rho_c)$ and so a higher critical density; similarly, the fraction of all SNe turning into SN Ia could also be smaller, pointing in the same direction.  This might be verifiable with a few well observed SN Ia.  

\begin{figure}[ht]
\vskip -36pt

\centering
\includegraphics[scale=1.05]{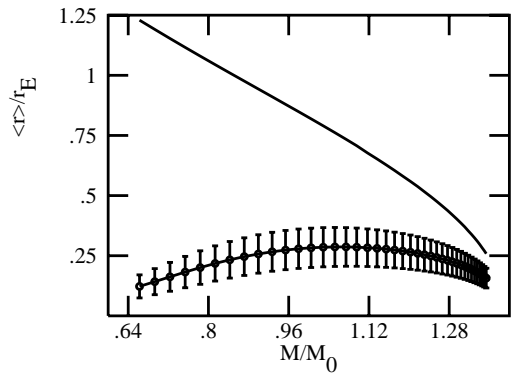}
\vskip -36pt

\caption{\baselineskip=18pt The mean nucleation point of the susy bubble relative to the stellar center with rms deviation
indicated based on a critical density of $10^7$ g/cc. For high mass stars the rms upper limit is
approximately the radius at which the density is equal to the critical density. 
The upper line gives the stellar radius as a function of white dwarf mass. $R_E$ is the Earth radius.}
\label{nucleationpoint} 
\end{figure}
\vskip 18pt
 
Although the most probable nucleation point of the susy core is at the stellar center, from Eq.\,(\ref{transprob2}) we can predict the mean distance from the stellar center of the nucleation (see fig.\,\ref{nucleationpoint}).  Thus the model
gives a clear quantitative prediction of the amount of ``off-centeredness" of the stellar explosion.  The fusion to $^{56}Ni$ is presumably centered in the
most dense part of the star outside the susy core regardless of where the susy energy release originates.

\subsection{Massive stars and the Black Hole Gap}

Applying the susy transition idea to massive stars as they agglomerate \cite{Sanders1970} within a dark matter clump in the very early universe also leads to a phase transition that may be fast enough to slow the collapse and eject much of the mass of the star before it ultimately collapses as a black hole (BH) at much lower mass.  This would then help to explain the BH mass distribution gap between about 30 and about a million solar masses, where we observe very few BHs.  Going through the numbers \cite{Chandra39} suggests $10^{5}$ solar masses as the upper limit for this mechanism since this is the dividing line to achieve white dwarf density before collapsing into a BH at higher masses.  We note that the general relativistic instability of
very massive stars found by Appenzeller and Fricke is at slightly higher mass \cite{Sanders1970}.

The maximum average density that a stellar conglomeration of mass $M$ can achieve before becoming a black hole is 

\be
    \rho_{max} = 1.83\,\cdot \,10^{6}\;\left(\frac{M}{10^5\,M_{\odot}}\right)^{-2}\,\mathrm{g/cc} \quad .
\label{rhoWD}
\ee

The peak density is expected to be $54$ times greater (Chandrasekhar 1939)

\be
    \rho_{peak} = 9.9\,\cdot 10^7\;\left(\frac{M}{10^5\,M_{\odot}}\right)^{-2}\,\mathrm{g/cc}\quad .
\ee

The numerical factor in Eq.\,(\ref{rhoWD}) is a typical white dwarf density $\rho_{\mathrm WD} = \frac{3\,M_0}{4\pi\,R_E^3}$.  Thus with a critical density in this range, the susy phase transition predicts an edge in the supermassive black hole distribution near $10^5\,M_{\odot}$.  Above this mass a conglomeration can become a black hole before the critical density for a phase transition is reached.  Below this mass the susy transition will take over and disrupt the collapse into a black hole.  The volume of the susy core which will eventually become a susy black hole is estimated \cite{growth} to be less than $0.1\%$ of the original stellar volume.  Thus the present model could help explain the gap in the black hole mass distribution \cite{Barth2005,Caramete}. 

This then leaves the range of about $10^{5}$ to about $10^{6}$ solar masses to make the first generation of super-massive BHs; these BHs can merge and then may help to explain the observed supermassive BH mass distribution \cite{Caramete}. 

\subsection{Supernova Characteristics}
Returning to supernova properties,
a clear prediction of our model is then the absolute rate of SN Ia, see Eq.\,(\ref{SNIarate}) and Eq.\,(\ref{SNIarateG}) in conjunction with fig.\,\ref{Rhogd} and the relative statistics in terms of mass, energetics, chemical composition, of white dwarfs which explode and with what delay.  A key prediction of our model is the shape of the delay time distribution, see fig.\,\ref{dndtaufig}.  Detailed observations and modelling should confirm that the total mass extends significantly below the Chandrasekhar limit, and that the original white dwarf was composed mostly of carbon and oxygen, since these are the stars in the preferred mass range.  A further prediction is the existence of very small mass compact susy remnants, observable in their passage through the interstellar medium 
due to radiation from accreting material. 
%

{\bf Acknowledgements}

Discussions with Ben Harms, Dean Townsley, and Jimmy Irwin greatly contributed to the development of the paper; discussions of PLB with Akos Bogdan, Alex Heger, Ed van den Heuvel, Norbert Langer, and Simon Portegies Zwart are gratefully acknowledged.  This research was supported in part by the DOE under grant DE-FG02-10ER41714.

\end{document}